\documentclass[aps,prl,twocolumn,groupedaddress,showkeys]{revtex4}
\usepackage{epsfig}

\begin{document}


\title{High temperature unfolding simulations of a single-stranded DNA i-motif}

\author{Jens Smiatek$^1$}
\author{Dongsheng Liu$^2$}
\author{Andreas Heuer$^1$}
\affiliation{$^1$Institute of Physical Chemistry, University of Muenster, D-48149 M{\"u}nster, Germany\\
	     $^2$Department of Chemistry, Tsinghua University, Beijing 100190, P. R. China}


\begin{abstract}
We present the results of high temperature 500 K Molecular Dynamics simulations of the DNA i-motif. The essential dynamics and the main unfolding pathways are compared to a biased metadynamics simulation at 300 K.
Our results indicate a remarkable agreement of the concerted motion at both temperatures. The transition can be described by a few number of eigenvectors indicating a simple unfolding mechanism.
Two main mechanisms for the unfolding pathway at 500 K can be detected which are in good agreement to the results of the biased simulation at 300 K.
\end{abstract}

\date{\today}

\maketitle
\section{Introduction}
\label{sec:introduction}
\begin{figure}[h!]
 \includegraphics[width=0.35\textwidth]{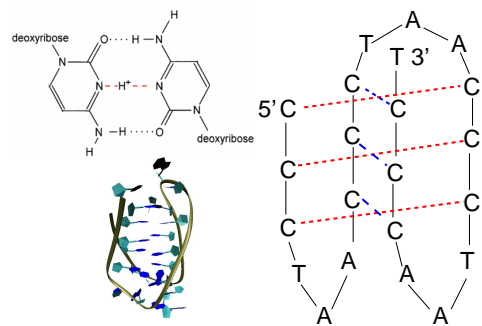}
 \caption{C-CH$^+$ pairing (bottom left) which is responsible for the formation of the DNA i-motif (top left) with the corresponding sequence (right) where C,T and A denote cytosine, adenine and thymine.
}		
\label{fig1}
\end{figure}

High temperature simulations are common tools for the investigation of protein unfolding properties \cite{Levitt93,Daggett02,Fersht03,Daggett04}. 
Due to the inherent long timescales apparent in the unfolding processes \cite{Daggett02,Karplus98}, 
a decrease in computation time is even nowadays still desirable. By the usage of elaborated temperatures which are typically above the protein melting temperature $T_M$, a 
significant decrease of unfolding times can be observed \cite{Levitt93,Daggett02,Fersht03,Daggett04}. As it is often assumed, an Arrhenius-type behavior  of unfolding kinetics is responsible for this remarkable
acceleration \cite{Wade07} while the pathway of unfolding is unchanged \cite{Daggett04}.\\   
Over the years, the desirability of using elaborated temperatures has lead to several novel insights.
It has been shown that the properties of the solvent related to the chosen water model at higher temperatures are in agreement to
experimental results \cite{Daggett02}. Hence the validity of the unfolding pathway is not influenced by spurious artefacts of the environment.
Furthermore it is known that at very high simulation temperature the unfolding process occurs more rapidly due to a simultaneous breaking of 
several intramolecular interactions such that intermediate states cannot be detected considerably \cite{Wade07}. To avoid this drawback, the concept of an upper critical 
temperature has been introduced \cite{Wade07}. Carrying out simulations below this temperature leads to an enhanced sampling of intermediate states such that the complete unfolding pathway can be studied in detail.
Hence the transformation of high temperature to low temperature pathways is methodologically validated although a detailed comparison is often missing.\\
The application of elaborated temperature unfolding simulations can be also useful to study the unfolding mechanisms for specific DNA and RNA structures due to large free energy barriers. 
Prominent examples for
these configurations are non Watson-Crick like structures in DNA like the G-quadruplex structures and the i-motif \cite{Gueron1993,Leroy2000,Trent2008,Liu2009,Mergny2002}.\\ 
The first one is formed by 
guanine (G) rich sequences \cite{Trent2008} while the latter is present in more cytosine (C) rich strands of DNA \cite{Leroy2000}. 
Although binding seems to be unfavorable for these structures, stability is achieved by a proton mediated cytosine binding between different strands or regions 
of the sequence resulting in a C-CH$^+$ pairing \cite{Gueron1993,Leroy2000,Liu2009,Mergny2002}. Hence an acidic environment is able to spend a proton which leads to a hemi-protonated cytosine mimicking an ordinary 
C-G binding as it is present in common DNA. It becomes clear that these structures have been found to be only stable at slightly acidic to neutral conditions resulting
in pH values ranging from 4.8 to 7.0 \cite{Gueron1993,Leroy2000,Tan2005}. Furthermore i-motif structures show a remarkable stability \cite{Tan2005} and have been found as tetrameric and double stranded complexes 
between different molecules 
although they are also occurring in single stranded DNA \cite{Leroy2000}. 
A picture of the C-CH$^+$ complex and the corresponding single stranded i-motif with its 
sequence is shown in 
Fig.~\ref{fig1}.\\
However, a detailed investigation of the function in the human cell is still missing although the applicability as a new class of possible targets for cancers and other diseases has been 
discussed \cite{Gupta1997,Hurley2001}. 
In contrast to this lack of knowledge, the usage of the i-motif in modern biotechnology has experienced an enormous growth over the last years \cite{Liu2009}.  
Since the i-motif becomes unstable at basic pH values, a systematic decrease and increase of protons in the solution by changing the pH value leads to a reversible folding and unfolding mechanism.
It has been shown that the unfolding and folding occurs on a timescale of seconds \cite{Liu2009,Tan2005}. Due to these properties the presence of i-motifs in modern nano applications is obvious.\\
Technological examples for the i-motif are given by molecular nanomachines \cite{Liu2009,Liu2003}, switchable nanocontainers \cite{Mao2007}, 
sensors to detect the pH value inside living cells \cite{Krishnan2009}, 
building materials for logic gate devices \cite{Qu2010} and sensors for distinguishing single walled and multi-walled carbon nanotube systems \cite{Qu2009}.\\
Recently it has been assumed \cite{Zhang2010}, that the grafting density massively influences the structure of an i-motif layer due to steric hindrance in the underlying folding and unfolding pathway.   
Hence a detailed investigation of the folding respectively unfolding pathway especially for high grafting densities in nanodevices of the i-motif and under physiological conditions is of prior importance.\\
In this paper we present the results of high temperature 500 K Molecular Dynamics simulations concerning the unfolding mechanism of a maximum unstable single stranded DNA i-motif structure without hemiprotonated cytosine. 
Our results indicate a fast initial decay of 
the i-motif leading to stable conformations. By detailed investigation of the essential dynamics of a biased metadynamics simulation, we were able to identify the main unfolding pathways at 300 K. The good agreement
to the high temperature unfolding pathway is validated. 
Our findings can be related to the calculated underlying free energy landscape which allows to identify the most stable conformations.\\
The paper is organized as follows. In the next section we present the theoretical background of the metadynamics method and the essential dynamics methodology. Then we introduce the numerical details and present
the results of our simulations. We conclude with a brief summary.
\section{Methods}
\subsection{Metadynamics}
The system we consider is described by a set of coordinates $S(x)$ evolving under the action of dynamics following the trajectory $S(x(t))$ and described by a canonical equilibrium distribution at temperature $T$. 
The set of
coordinates $S(x)$ may include atomic positions or angles as well as any other auxiliary collective variable representing the characteristics of the system.\\
If the system shows metastability, some regions separated by large energy barriers cannot be explored by the evolution of the trajectories for low temperatures in a reasonable simulation time.
Thus reflecting the original idea of metadynamics, an additional potential energy at specific constant times $t_1, t_2,...t_{_{N}}$ is applied on the
trajectory $x(t)$ to overcome the barriers and to accelerate the rare events \cite{Laio02}. 
Typically this potential energy is given by Gaussian hills which are summed at time $t$ to
\begin{equation}
V_{_{G}}(S(x),t)=\omega \sum_{t^{\prime} = t_{_{1}},t_{_{2}},...}\exp\left(-\sum_{_{\alpha}}^d\frac{(S_{_{\alpha}}(x)-s_{_{\alpha}}(t^{\prime}))^2}{2\delta s^2}\right)
\label{eq:mtd}
\end{equation}
where $s_{_{\alpha}}(t^{\prime})=S_{_{\alpha}}(x(t))$ are the $d$ collective variables, 
$\omega$ denotes the Gaussian height and $\delta s$ is the Gaussian width. By applying this procedure subsequently, free energy minima can be escaped and the unfolding pathway can be 
identified \cite{Laio08,Ensing05,Ensing06}. 
Furthermore it has been shown \cite{Laio08} that the added potential energy resembles the underlying free energy
\begin{equation}
\lim_{_{t\rightarrow\infty}}V_{_{G}}(s,t) \sim -F(s)
\end{equation}
after a long simulation time.
To overcome certain drawbacks \cite{Laio08,SmiatekWHM}, we use a recently published variant of the metadynamics method \cite{SmiatekWHM}. 
Within this method, the metadynamics potential is calculated on a grid leading to 
a constant calculation time in contrast to ordinary metadynamics. The refinement of the resulting landscape is achieved by a histogram reweighting procedure of several biased simulation runs \cite{SmiatekWHM,Kumar92}.
A projection scheme allows to change the collective variables a posteriori \cite{SmiatekWHM}.
\subsection{Essential Dynamics}
The main part of a successful calculation relies on the choice of appropriate collective variables on which the free energy landscape and the resulting unfolding pathway is spanned.
Well-suited collective variables are the essential eigenvectors of the system
\cite{Amadei93}. It has been recently demonstrated that the application of eigenvectors as collective variables results in adequate descriptions of the free energy landscapes in ordinary Metadynamics 
computations \cite{Spiwok07,Spiwok08}.
Thus we follow this approach due to the assumption that nearly all relevant motion is captured in the
first eigenvectors of the system \cite{Amadei93}. 
In the following we give a brief description of the method.\\
The essential eigenvectors can be calculated by the superimposed coordinates $\vec{r}$ of $N$ atoms of the system which build the covariance matrix ${\bf C}$ via
\begin{equation}
{\bf C}=<(\vec{r}-<\vec{r}>)(\vec{r}-<\vec{r}>)^T>
\end{equation}
where $i,j = 1,2,\dots 3N$ and the brackets $<\ldots>$ denote the reference value. The diagonalization of ${\bf C}$ leads to
\begin{equation}
\bf{C}={\bf E}\Lambda{\bf E}^{-1}
\end{equation}
where ${\bf E}$ is a matrix of eigenvectors and ${\bf \Lambda}$ is a matrix of eigenvalues marking the positional fluctuations.
Sorting the eigenvalues in decreasing order allows to identify the largest positional fluctuations by the first eigenvectors which form the essential subspace with all important structural transitions \cite{Amadei93}.
The projection at time $t$ on the $i-$th eigenvector $\vec{e}_i$ is then defined by
\begin{equation}
p_i(t)=(\vec{r}(t)-<\vec{r}>)\cdot \vec{e}_i
\end{equation}
with the specific eigenvectors ranging from $i=1,2\dots 3N$ \cite{Spiwok07,Spiwok08}.\\
To compare the motion of several subsets of eigenvectors for different trajectories, a class of functions has been introduced \cite{Berendsen96,Amadei96,Nola99, Grubmuller2001}.
The surface overlap $\mathcal{O}_{_{NM}}$ between $N$ eigenvectors $\vec{\mu}_i$ from one set and $M$ eigenvectors from another subset $\vec{\nu}_j$ can be calculated by the squared inner product
\begin{equation}
\mathcal{O}_{_{NM}} = \frac{1}{N}\sum_{_i}^N\sum_{_{j}}^M(\vec{\mu}_i\vec{\nu}_j)^2
\label{eq:solp}
\end{equation}
which gives an estimate for the similarity of two sets. If the spaces are completely orthogonal the relation gives $\mathcal{O}_{_{NM}}=0$ in contrast to identity which results in $\mathcal{O}_{_{NM}}=1$.
Additionally the inner product matrix can be calculated by
\begin{equation}
o_{_{ij}} = \sqrt{(\vec{\mu}_i\vec{\nu}_j)^2}
\end{equation} 
which allows to derive a detailed comparison between the motion of different subset or trajectories and the corresponding eigenvectors.
\section{Simulation details}
We have performed our Molecular Dynamics simulations of the i-motif in explicit TIP3P solvent at T=300 K and T=500 K by the GROMACS software package \cite{Gromacs}.
The single DNA strand consists of 22 nucleic acid bases given by the sequence $5^{\prime}-CCC-[TAA-CCC]_3-T-3^{\prime}$ where $T$, $A$ and $C$ denote thymine, adenine and cytosine. 
We modeled this structure which is directly related to the sequence
used in \cite{Liu2003} 
by slight modifications of the PDB entry 1ELN \cite{PDB}.
Our periodic simulation box contains 5495 TIP3P water molecules and 22 sodium ions to compensate the charging of the DNA. All interactions have been calculated by using the ffAMBER03 force field \cite{Sorin2005}.
After energy minimization, the initial
warm up phase of 1 ns has been performed by keeping the position of the i-motif restrained. 
The cubic simulation box with periodic boundary conditions has a dimension of $(5.41\times 5.41\times 5.41)$ nm. 
We applied a Nose-Hoover thermostat to the system 
where all bonds have been constrained by the LINCS algorithm \cite{Gromacs}. Electrostatics have been calculated by the PME algorithm \cite{Gromacs} and the time step was $2$ fs. 
The calculation of the free energy landscape has been performed by the method presented in \cite{SmiatekWHM,SmiatekDNA}. The biased metadynamics simulations at 300 K for the investigation of the unfolding pathway 
have been conducted by the program plug-in PLUMED \cite{Plumed}. The Gaussian hills were set each 2 ps with a height of 0.25 kJ/mol and a width of 0.25 nm. The corresponding reaction coordinates for 
the biased energy are the distance between nucleobases C1 and T22 and the distance between the center-of-mass for the C1-T22 to the A11 nucleobase as they were also used in \cite{SmiatekDNA}. 
The final eigenvector free energy landscapes
have been calculated by the projection scheme proposed in \cite{SmiatekWHM} where the landscape has been refined via 15 biased simulations of 10 ns at 300 K and 500 K. A detailed protocol for the explicit 
calculation of the eigenvector free energy landscape in correspondence to the projection scheme is presented in \cite{SmiatekWHM,SmiatekDNA}.\\  
For a detailed investigation of the unfolding mechanism we furthermore conducted five 500 K simulations and five 300 K unbiased simulations each with 10 ns duration and averaged the resulting values. 
The root-mean square deviation is used for the study of the unfolding pathway
\begin{equation}
RMSD(t,t_0)=\sqrt{\frac{1}{N^2}\sum_{i}^N\sum_j^N(\vec{r}_{ij}(t)-<\vec{r}_{ij}(t_0)>)^2}
\end{equation}
with the positions $\vec{r}_i$ of $N$ atoms for different times and the distances $r_{ij}$ between two atoms  
where $<\vec{r}_i(t_0)>$ is the reference or average position. 
Due to reasons of simplicity, we paired each three nucleobases in one group resulting in the 
sequence CYS1-TAA1-CYS2-TAA2-CYS3-TAA3-CYS4-T.  
We chose the initial structure as reference position and calculated the RMSD
for the 500 K and the 300 K unbiased simulations for several groups.
\section{Numerical results}
\begin{figure}[h!]
 \includegraphics[width=0.5\textwidth]{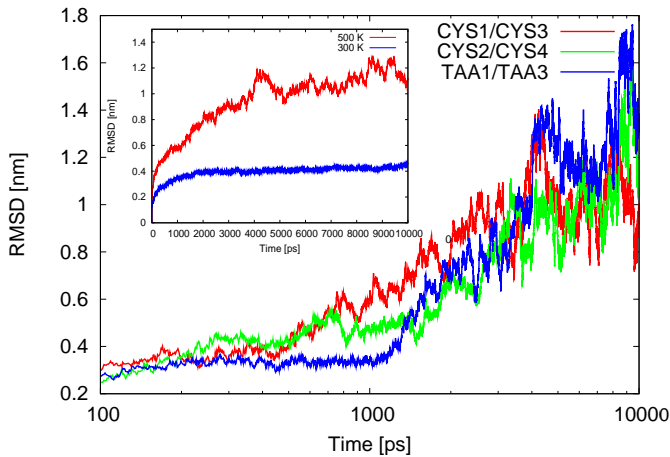}
 \caption{Root-mean-square deviations for different nucleobase combinations in the 500 K simulation and for the complete molecule at 300 K and 500 K (inset). 
}		
\label{fig2}
\end{figure}

We start this section by presenting the values for the root-mean square deviation of the unbiased simulations at 500 K and 300 K in Fig.~\ref{fig2}.
The initial opening of the i-motif appears on a timescale up to 200 ps for the 500 K simulations. This is obvious by regarding the results for the slightly increasing RMSD until it becomes nearly constant 
from 200 ps to 400 ps. 
After 400 ps a stable structure can be 
observed with minor fluctuations indicated by a constant RMSD of the TAA1/TAA3 which is present up to 1.2 ns. As it has been discussed in \cite{SmiatekDNA} for 300 K, a reason for this stable conformation is a constant 
number of hydrogen bonds between the TAA1/TAA3 group.\\
Significant fluctuations in the CYS1/CYS3 and the CYS2/CYS4 group after 600 ps can be observed which may be responsible for the opening of the TAA1/TAA3 group on a timescale up to 3 ns. 
Finally on longer timescales than 3 ns the structure unfolds into a fully extended strand in the
500 K simulations as it can be seen by the large RMSD of all groups. 
It is remarkable that the 300 K simulations do not reach the fully unfolded structure as the inset of Fig.~\ref{fig2} indicates. 
Hence it becomes clear that the usage of biasing methods at these lower temperatures
is required.\\ 
A more detailed analysis of the unfolding pathway can be performed by the investigation of the essential dynamics \cite{Amadei93}.
To study the concerted motion of the molecule and to indicate the main structural transitions we analyzed the essential eigenvectors of the system.
\begin{figure}[h!]
 \includegraphics[width=0.5\textwidth]{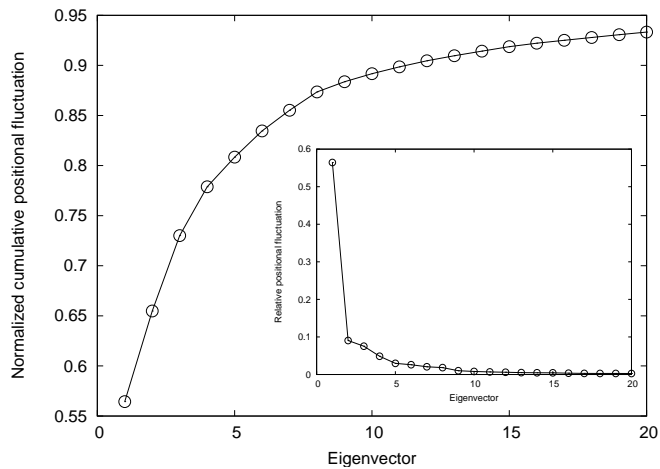}
 \caption{Normalized cumulative positional fluctuation of the eigenvectors for the 500 K simulations and the corresponding relative positional fluctuation (inset).
}		
\label{fig3}
\end{figure}   
Most of the motion at 500 K which is about 66\% is expressed in the first two eigenvectors as Fig.~\ref{fig3} indicates. 
Nearly 57\% of the overall motion is incorporated in eigenvector 1. The essential subspace is formed by 
the first eight eigenvectors with roughly 88\%. 
The relative positional fluctuations again indicate that eigenvector 1 is dominating the overall motion with significant contributions from eigenvectors 2 and 3. 
The presence of all eigenvectors larger than 8 is negligible. It can be concluded that this represents an unfolding pathway which can be effectively 
described by a small number of  
eigenvectors.\\
The motion of the first two eigenvectors at 500 K is presented in Fig.~\ref{fig4}. Eigenvector 1 mainly describes the opening of the end-to-end 
distance by a stretching mode. Eigenvector 2 can be identified as the initial opening of the i-motif by the relaxation into a planar structure.\\ 
\begin{figure}[h!]
 \includegraphics[width=0.5\textwidth]{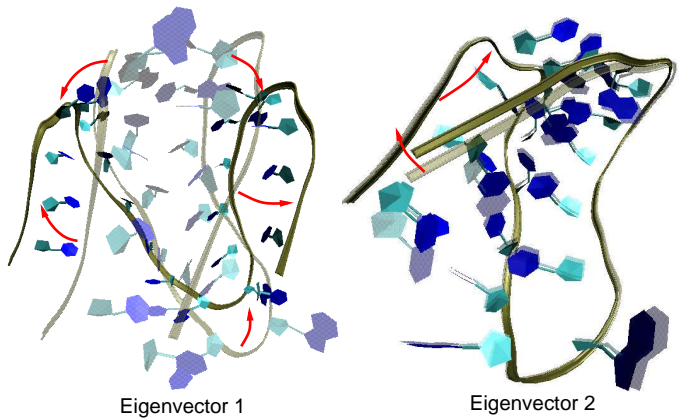}
 \caption{The motion of the eigenvectors 1 and 2 at 500 K. The arrows illustrate the movement along the eigenvectors.}		
\label{fig4}
\end{figure}
To compare the unfolding pathways of 500 K to lower temperatures, we performed a 50 ns metadynamics simulation at 300 K. This 
allows to overcome energetic barriers and to identify the main unfolding pathways which can be compared to the high temperature simulations by an analysis of 
the essential dynamics.\\
For this we investigated the eigenvectors for the biased metadynamics run at 300 K, the unbiased 500 K and 300 K simulations.
The projection is performed along the unbiased 500 K subset of eigenvectors 1 and 2 on a timescale of 10 ns.\\ 
\begin{figure}[h!]
 \includegraphics[width=0.5\textwidth]{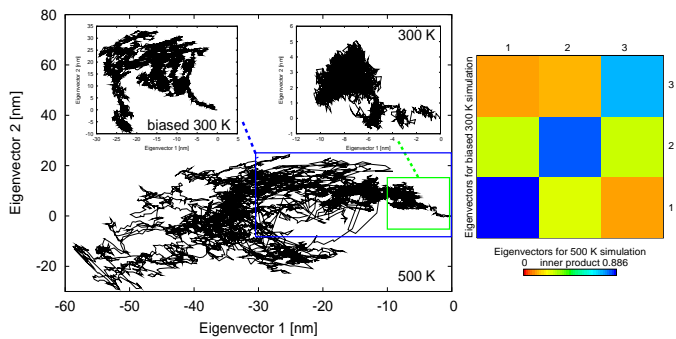}
 \caption{{\bf Left:} Unfolding pathways for the eigenvectors 1 and 2 after 10 ns for the unbiased 500 K simulations and the biased 300 K simulations (top left) and the unbiased 300 K simulations (top right). {\bf Right:}
Inner product matrix $o_{_{ij}}$ for the biased 300 K simulations and the unbiased 500 K simulations. 
}
\label{fig5}
\end{figure}		
Fig.~\ref{fig5} illustrates the importance of elaborated temperatures respectively the usage of biasing potentials to discover the unfolding pathway.
It can be seen that the biased simulations at 300 K resembles the high temperature trajectory in good agreement although it is limited to a smaller region. 
This is also true for the unbiased 300 K simulation which is 
restricted to a very small subspace due to occurring trapping in a specific configuration whereas the main unfolding pathway is identical. 
The results indicate that the unfolding mechanism is remarkably accelerated by elaborated temperatures or biasing methods and obviously resembles the real low temperature unfolding pathway.\\
The results above are additionally validated by the calculation of 
the squared inner product $\mathcal{O}_{_{ij}}$ (Eqn.~\ref{eq:solp}) for the complete biased 300 K and the unbiased 500 K simulations and the eigenvector inner product $o_{_{ij}}$ which is presented on the right 
side of Fig.~\ref{fig5}. 
The squared inner product for the first three eigenvectors is $\mathcal{O}_{_{33}}=0.787$ with a square root of $0.887$ which indicates a nearly identical subspace of the first three eigenvectors at different 
temperatures.   
These values are supported by the inner product matrix $o_{_{ij}}$. 
Eigenvector 1 has an inner product of $0.886$ indicating a remarkable identity of both trajectories. This is also true for 
eigenvectors 2 and 3 with values of $0.841$ and $0.753$. Hence the overall motion of the first three eigenvectors which describe 73 \% of the concerted motion (Fig.~\ref{fig4}) is directly comparable
at both temperatures.
This proves the assumption that the usage of elaborated temperatures results in unfolding pathways that are comparable to those derived at lower temperatures.\\
\begin{figure}[h!]
 \includegraphics[width=0.5\textwidth]{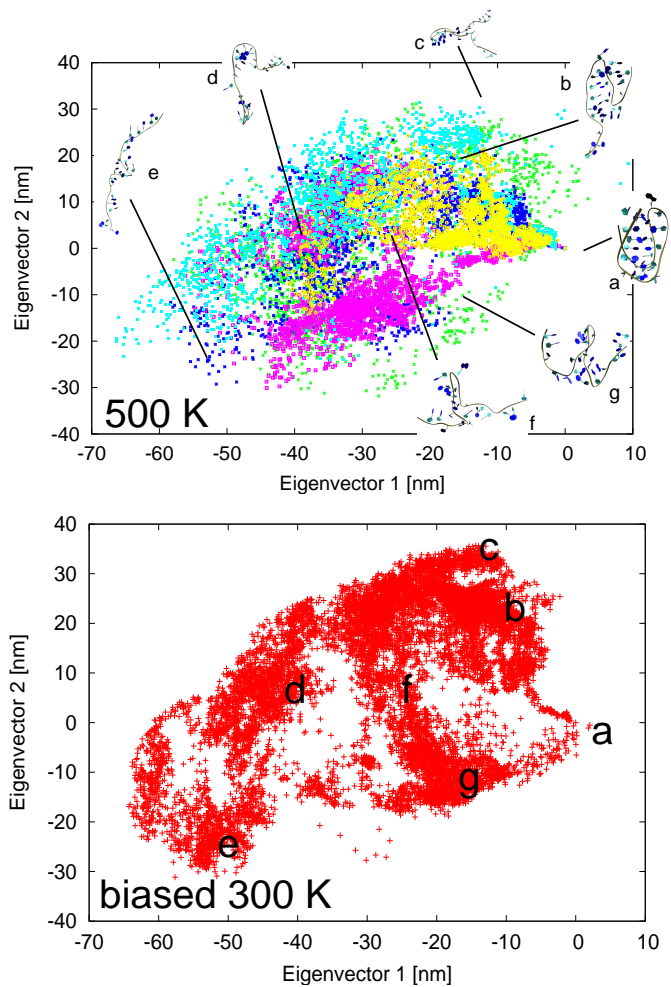}
 \caption{Explicit trajectories for all 500 K simulations (top) and the biased 300 K (bottom) simulation. Typical conformations at certain regions and both temperatures are shown as snapshots and denoted by letters.
}		
\label{fig6}
\end{figure}
\begin{figure}[h!]
 \includegraphics[width=0.5\textwidth]{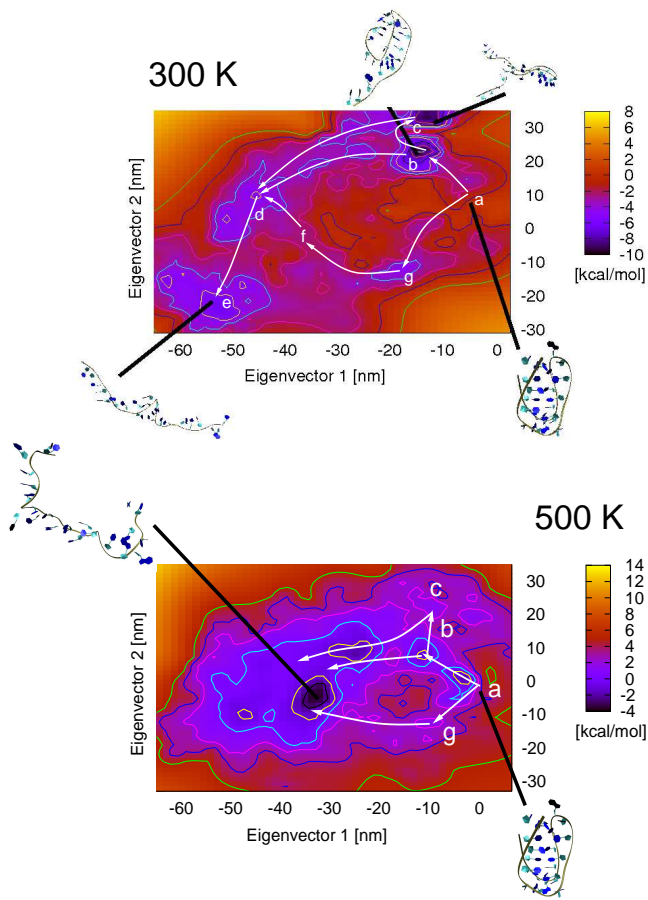}
 \caption{Free energy landscape at 300 K (top) and 500 K (bottom) for the eigenvectors 1 and 2 with the corresponding determined unfolding pathways. The lines correspond to free energy differences of 1.5 kcal/mol.
}
\label{fig7}
\end{figure}
The results for all unbiased 500 K simulations and the complete biased 300 K simulation trajectory are presented in Fig.~\ref{fig6}. 
It is obvious that similar regions are explored at both temperatures and the overall displacement at both temperatures is comparable.
Several conformations can be identified along the trajectories starting with the initial i-motif. For a detailed analysis of the stability of the presented conformations and the unfolding pathway, the underlying 
free energy landscapes have to be calculated. This allows to decide whether a conformation is stable or represents a global minimum which can be directly related to the detected unfolding path. 
The free energy landscapes at 300 K and 500 K are presented in Fig.~\ref{fig7}.\\ 
It is obvious that the most stable conformation is given by a stretched structure at 500 K. This is in contrast to 300 K 
where it can be seen that two stable hairpin conformations (b) and (c) represent the lowest free energy conformations. 
It can be assumed that the local ordering of water molecules is mainly responsible for this effect due to the fact that it is largely dominated by electrostatic interactions of the DNA. 
At larger temperatures, the dominance of the Coulomb interactions in comparison to 
thermal energy becomes smaller such that additional solvent and molecular configurations are accessible. Finally this results in a change of the entropy of these configurations such that the free energy minima are adjusted
by entropic contributions.
Although the minima and the barriers differ in their height and their relative position, the overall shape
of the landscapes is similar. Entropic contributions as discussed above may be responsible for these slight deviations. 
The good agreement becomes also obvious by studying the unfolding pathways and the corresponding free energies. 
Thus it can be concluded that comparable unfolding pathways can be observed as it was also discussed above. However, the realization of the unfolding pathway in an unbiased simulation at lower 
temperatures is dependent on the energy barriers that are occurring on the pathway. As Fig.~\ref{fig7} indicates, the general movement of the unfolding process is comparable at both temperatures.\\
The hairpin conformations (b) and (c) at 300 K belong to free energy minima and are the reason for the stagnating root-mean square deviation presented in Fig.~\ref{fig2}. 
Hence the overcome of the energetic barriers for these structures is cumbersome in a 300 K unbiased simulation. Thus it can be concluded that the 
unfolding pathway at lower temperatures is mainly realized by a transition into conformations (b) and (c). In contrast to 300 K, the free energy landscape at 500 K  
indicates these conformations as local minima 
with low barriers such that the transition into a fully extended strand representing the global energy minimum can be realized.\\  
Starting with the i-motif (a) the initial opening of one end with the groups
CYS3-T is sufficient for the transition in the global minimum (b) at 300 K. 
Overcoming this stable configuration results in two possible pathways. The first possibility is the transition into the neighbored minimum which is given by a planar hairpin structure (c). 
The other pathway follows the opening of the strand with groups CYS1 and TAA1 into a planar s-shaped structure (d). 
The final transition is then realized by an unfolding into the fully extended strand (e). 
However, the last pathway is better defined for 300 K due to the fact that the completely unfolded region is energetically more favorable for the 
500 K simulations as the free energy landscapes indicate. However all configurations have been clearly visited in the 500 K simulations.\\ 
A further pathway is the unfolding into a twisted structure (g) 
where the strand with groups CYS1-T follows a rotational motion. 
After this move, the ends of the strand relax into structure (f). By a second 
process, the twisted configuration transforms into conformation (d) which is also present in the first unfolding pathway at 300 K. 
The final transition into the stretched structure (e) is identical to the first unfolding 
pathway. At 500 K this pathway visits conformations (g) until the configuration fully unfolds on different pathways as discussed above.  
\section{Summary and Conclusion}
We have simulated the unfolding of the DNA i-motif via Molecular Dynamics simulations in explicit solvent at 500 K and at 300 K.  Our results indicate that the initial structure of the i-motif 
vanishes on a timescale of a few nanoseconds. Thus our results are in agreement to experimental observations  
which indicated that an unprotonated i-motif is not stable at room temperature \cite{Leroy2000,Tan2005,Liu2009}.\\ 
We have shown that unfolding pathways can be investigated either by the usage of elaborated temperatures as well as biasing potentials. The occurring trajectories are similar as a detailed analysis of the essential dynamics 
indicated. Hence we have presented a method to prove the transformation of high temperature unfolding trajectories onto lower temperatures.  
Due to this similarity we can identify two main unfolding pathways for the DNA i-motif at both temperatures. 
The most relevant one at 300 K visits stable hairpin configurations from which it unfolds only due to thermal activation or the presence of biasing potentials 
into an extended structure in a reasonable simulation time. 
The other one includes
torsional motion and is energetically less favorable.\\
The calculated free energy landscapes at 500 K and 300 K 
are similar in their global shape which leads to comparable unfolding pathways. Hence it can be concluded that this property is the main reason for the success of high temperature
simulations. Finally our results indicate that the analysis of the unfolding pathways is only reasonable in combination with the calculated free energy landscapes. This becomes specifically important by comparing the 
global energetic minima which may result in only partly visited unfolding pathways. 
Hence it can be concluded that at lower temperatures the i-motif mainly unfolds 
into the discussed hairpin configurations. The fully extended strand represents the global energy minimum only at higher temperatures. Finally it is obvious that
the detailed unfolding pathway is better defined for 300 K although the main configurations are visited at both temperatures.\\
As it has been discussed, the determination of the unfolding pathway of the i-motif is of prior importance regarding the applicability in nanomachines \cite{Liu2003,Mao2007,Zhang2010,Liu2009}.
Hence our results allow to improve the applicability of i-motifs as functional materials in nanotechnology due to the explicit knowledge of the unfolding pathway and its stable conformations. 
Finally the investigation of the unfolding properties may also lead to a more detailed knowledge about the molecular mechanisms in the human cell. This is in particular interesting due to the fact that the knowledge 
of the function of the i-motif is still missing.
\section{Acknowledgments}
We thank the Deutsche Forschungsgemeinschaft (DFG) through the transregional collaborative research center
TRR 61 for financial funding. 

\end{document}